# 17 GHz Lossless InP-Membrane Active Metasurface

Taichiro Fukui[1,], Kei Sumita[1], Hiroki Miyano[1], Go Soma[1], Warakorn Yanwachirakul[1,†], Eisaku Kato[1], Ryota Tanomura[1], Jiahao Liu[1], Toshiki Yamada[2], Akira Otomo[2], Kasidit Toprasertpong[1], Mitsuru Takenaka[1], Shinichi Takagi[1], Yoshiaki Nakano[1], and Takuo Tanemura[1,*]

[1]Department of Electrical Engineering and Information Systems, The University of Tokyo, 7-3-1 Hongo, Bunkyo-ku, Tokyo 113-8656, Japan
[2]National Institute of Information and Communications Technology, 588-2 Iwaoka, Nishi-ku, Kobe, Hyogo 651-2492, Japan
[†]Present address: Department of Physics, Faculty of Science, Chulalongkorn University, 254 Phayathai Road, Patumwan, Bangkok, Thailand, 10330.
[*]fukui@hotaka.t.u-tokyo.ac.jp; tanemura@ee.t.u-tokyo.ac.jp

## ABSTRACT

High-speed active metasurfaces enable spatiotemporal control of incident light within an ultra-thin layer, offering new possibilities for optical communication, computing, and sensing. However, a fundamental tradeoff between electrical conductivity and optical absorption of the material has hindered the realization of active metasurfaces that simultaneously achieve broad modulation bandwidth and low optical loss. Here, we experimentally demonstrate a high-speed active metasurface operating in the 1.5-$\mu$m wavelength range that realizes a record-high modulation bandwidth of 17.5 GHz, while maintaining a high quality ($Q$) factor of 102 and an ultra-low optical loss of 0.56 dB. The key enabling technology is the indium-phosphide (InP) membrane platform; an n-type InP offers both high electron mobility and low free-carrier optical absorption, making it an ideal material for active metasurface devices. The high-$Q$ Friedrich-Wintgen quasi-bound-states-in-the-continuum mode inside the InP-membrane high-contrast grating (HCG) is utilized to trap the normally incident light within an organic electro-optic (OEO) material, enabling efficient modulation. InP HCG also serves as an ultralow-resistance interdigitated electrodes for applying high-speed electrical signals to the OEO material, thereby offering 50-fold improvement in modulation bandwidth compared to conventional silicon-based counterparts. Our work paves the way towards high-speed, low-loss active metasurfaces for spatiotemporal control of light beyond the gigahertz regime.

## Introduction

Metasurfaces are flat optical devices composed of nanostructures that enable complex manipulation of lightwaves within an ultra-thin device. Over the past decade, various passive metasurface devices, such as flat metalenses[1–3], polarimeters[4–6], optical hybrids[7–9], and color routers[10–12], have been demonstrated, achieving the functionalities impossible with conventional optics[13–15]. To further extend the capabilities of metasurfaces to the spatiotemporal regime, development of reconfigurable metasurfaces, also known as active metasurfaces, is desired[16–21]. Of particular interest is the realization of ultra-high-speed active metasurfaces surpassing the gigahertz (GHz) range, which hold immense potential to revolutionize diverse fields of optics, including large-capacity optical communication[22–24], optical computing[25–28], and high-speed optical sensing[29–31].

Active metasurfaces can be realized either by mechanically displacing the photonic nanostructures[31–34] or by modulating the refractive index of materials[35–60]. To realize a high-speed device operating beyond the GHz regime, the latter strategy is required, utilizing the high-speed electro-optic (EO) effects such as the Pockels effect[38–55], free carrier effect[56–59], and quantum-confined Stark effect[60]. However, such high-speed EO effects typically fall short of effectively modulating the incident light within the limited interaction length of a thin metasurface.

Therefore, optical resonating structures are typically utilized in the EO-effect-based active metasurfaces, where the incident light is trapped within a thin metasurface layer to enhance the light-matter interaction. The quality ($Q$) factor of the resonator should not be so high that it restricts the modulation bandwidth, but needs to be sufficiently large ($Q > 100$) to achieve adequate modulation efficiency. Then, to efficiently apply high-speed electrical signals to the EO material, it is crucial to integrate highly conductive electrodes at the vicinity or within the optical resonating mode to enable efficient modulation. At the same time, they must not introduce optical loss to the resonator. A lossy cavity not only degrades the $Q$ factor, but also severely restricts the wavefront modulation capability; a lossy cavity generally becomes under-coupled, whereas efficient phase modulation with $2\pi$ tunable range requires an over-coupled cavity[61].

Nevertheless, simultaneously satisfying these two requirements is challenging. Metals can serve as ideal low-resistance electrodes, but they suffer from a plasmonic loss, which inevitably reduces the $Q$ factor of the cavity. On the other hand, doped silicon (Si) has been widely employed in the previous work to achieve high $Q$ resonance[46–50,56–59]. However, Si also suffers from the strict tradeoff between the high electrical conductivity and low optical loss; while it is essential to increase the doping concentration to reduce the electrical resistance, it simultaneously leads to substantial increase in optical absorption through the free carrier plasma effect of Si. As a result, the realization of a low-loss active metasurface with > GHz EO bandwidth remains elusive to date.

Here, we demonstrate a novel platform that comprises an indium-phosphide (InP) membrane high-contrast grating (IM-HCG) embedded with organic electro-optic (OEO) material to realize ultra-high-speed active metasurfaces, operating at 1.5-µm wavelength range (Figs. 1a-d). As shown in Figs. 1e-g, n-InP outperforms n-Si both in terms of electron mobility and optical absorption, thereby offering substantial improvements over Si-based active metasurfaces. Using n-InP for both the highly conductive electrode and low-loss HCG-based optical resonator, our device efficiently traps the incident light within the OEO material and modulated by applying electric field via InP. The IM-HCG embedded with EO polymer is successfully fabricated to demonstrate a record-high 3-dB modulation bandwidth of 17.5 GHz, optical resonance with a moderately high $Q$ factor of 102, and ultra-low insertion loss of 0.56 dB. The measured modulation efficiency of the resonant wavelength is 18 pm/V, which can be enhanced by employing an OEO material with higher EO coefficient and by improving the fabrication process.

## Results

### Concept and operating principle of IM-HCG active metasurface

Figures 1a-c show the schematic of the proposed IM-HCG active metasurface. The IM-HCG is composed of arrayed InP membrane bars on a quartz (SiO$_2$) substrate, which are embedded with an OEO material. An HCG generally exhibits a wide variety of optical characteristics by judiciously designing its thickness $d$, period $\Lambda$, and bar width $w$[46,62] (see Supplementary Fig. S1). Notably, at the dual-transverse-mode region ($1.27 < \lambda/\Lambda < 2.2$ in supplementary Fig. S1), high-$Q$ resonances can be obtained at specific values of $d$ through the bimodal resonance effect[46,62,63], which are also recognized as Friedrich-Wintgen quasi-bound-states-in-the-continuum (FW-qBICs)[64]. In particular, a high-$Q$ resonating mode is obtained when $d/\Lambda \sim 0.8$ (Supplementary Fig. S1 inset). The electric field intensity profile of this mode for $\lambda = 1550$ nm is depicted in Fig. 1d. The field is strongly confined inside the narrow gap of HCG, enabling efficient modulation of light through the refractive index change of the OEO material.

The InP bars serve not only as an optical resonator but also as interdigitated electrodes for applying electric field to the OEO material (Fig. 1b), so that its refractive index is modulated through the Pockels effect[46,47]. As a result of the resonance wavelength shift, the reflectance and transmittance of the active metasurface are complementarily modulated as depicted in Fig. 1a inset. The proposed device is lossless if the material absorption is negligible. In such a case, the sum of the reflectance and the transmittance becomes unity; the device functions as a unitary beam splitter, whose splitting ratio can be modulated at a high speed.

The modulation bandwidth of the device is determined by the RC time constant of the InP bars (Fig. 1c). Therefore, it is crucial to reduce the electrical resistance of the InP bars, typically by increasing the doping concentration $N_D$. However, higher doping concentration generally induces free carrier absorption. Thus, the doping concentration needs to be kept below a certain level where the effect of optical absorption is negligible. To this end, n-InP is superior to n-Si due to its excellent characteristics in terms of optical absorption and electrical conductivity[65–67]. First, the electron mobility $\mu$ of n-InP outperforms that of n-Si by nearly an order of magnitude at the same doping concentration (Fig. 1e). Second, the doping-induced optical absorption of n-InP is one order smaller than that of n-Si (Fig. 1f). Under the same constraint on the maximum allowed optical absorption, therefore, significantly larger doping concentration can be employed for n-InP compared to n-Si. As a result, the resistivity of n-InP bars, which is inversely proportional to $\mu N_D$, can be one to two orders of magnitude smaller than that in the case of n-Si (Fig. 1g). Indeed, for an absorption coefficient of $\alpha = 10$ cm$^{-1}$, the resistivity of n-InP is 50-fold smaller than n-Si.

### Numerical analysis

Figure 2a shows the numerically simulated reflectance spectrum of IM-HCG ($\Lambda = 760$ nm, $d = 630$ nm, and $w = 380$ nm) with various doping concentrations $N_D$ (see the **Methods** for details). Even at a large doping concentration of $N_D = 1 \times 10^{19}$ cm$^{-3}$, the reflectance peak at the resonant wavelength is kept high above 95%. For comparison, Fig. 2b shows the case for n-doped Si-membrane (SM)-HCG ($\Lambda = 748$ nm, $d = 540$ nm, $w = 374$ nm) for various values of $N_D$. Unlike IM-HCG, the reflectance peak drops rapidly as $N_D$ exceeds $1 \times 10^{18}$ cm$^{-3}$. These results directly reflect the one-order-of-magnitude larger free-carrier absorption of Si compared with InP (Fig. 1f).

Figure 2c summarizes the impact of the doping concentration on the $Q$ factor as well as overall optical loss of the modulator for both cases. Here, the optical loss (dB) is defined as $-10\log_{10}(R+T)$, where $R$ and $T$ denote the reflectance and transmittance at the resonant wavelength, respectively. For IM-HCG, the $Q$ factor is larger than 900 and the optical loss is as small as 0.24 dB even at a relatively high doping concentration of $N_D = 1 \times 10^{19}$ cm$^{-3}$, thanks to the small free carrier absorption of n-InP. In contrast, the $Q$ factor degrades and the optical loss increases rapidly at $N_D > 1 \times 10^{18}$ cm$^{-3}$ for SM-HCG.

Figures 2d and 2e show how the reflectance spectrum changes as we modulate the refractive index of the OEO material $n_{EO}$ for the IM-HCG and SM-HCG, respectively. We assume $N_D = 1 \times 10^{19}$ cm$^{-3}$ in both cases. A large resonant wavelength shift of $\Delta\lambda/\Delta n_{EO} \simeq 450$ nm/RI is obtained for both IM-HCG and SM-HCG, thanks to the strong confinement of light within the OEO material. Figure 2f compares the characteristics of the IM-HCG and SM-HCG when they are used as reflective optical intensity modulators. The operating wavelength is selected to be 1547.4 nm for IM-HCG and 1548.6 nm for SM-HCG. A small insertion loss of 0.21 dB is obtained with IM-HCG, while that of SM-HCG is as large as 4.3 dB. Moreover, thanks to the large $Q$ factor of IM-HCG, 5-dB modulation can be obtained at a small index change of $\Delta n_{EO} = 2.2 \times 10^{-3}$. This corresponds to a required driving voltage of $V_{pp} = 5.6$ V, assuming a simple parallel-plate model and the EO coefficient of the state-of-the-art OEO materials ($r_{33} = 200$ pm/V)[68–70]. In contrast, larger index change of $\Delta n_{EO} = 3.5 \times 10^{-3}$ is required to achieve 5-dB modulation for SM-HCG due to its degraded $Q$ factor. This corresponds to a required voltage of $V_{pp} = 8.7$ V.

Figure 2g shows the simulated 3-dB EO bandwidth as a function of the doping concentration for IM-HCG and SM-HCG. When the device size is $40 \times 40$ µm$^2$ and $N_D = 1 \times 10^{19}$ cm$^{-3}$, the 3-dB bandwidth of IM-HCG active metasurface can exceed 40 GHz, which is comparable to that of waveguide-based EO modulators deployed in current optical communication systems. On the other hand, the 3-dB bandwidth of SM-HCG is limited to around 3 GHz at $N_D = 1 \times 10^{19}$ cm$^{-3}$. Moreover, at such a large $N_D$, the optical absorption is non-negligible and exceeds 3 dB for SM-HCG (Fig. 2c). At $N_D = 1 \times 10^{18}$ cm$^{-3}$, where the optical absorption is less than 0.34 dB and comparable to that of IM-HCG at $N_D = 1 \times 10^{19}$ cm$^{-3}$, the 3-dB bandwidth is further reduced to 800 MHz. Thus, we can achieve 50-fold improvement in the EO bandwidth by using IM-HCG, reaching tens of gigahertz, without increasing the insertion loss.

### Device fabrication and characterization

Figure 3 shows the IM-HCG device, which was fabricated by bonding a 590-nm-thick n-InP membrane layer on a quartz (SiO$_2$) substrate (see **Methods** for details of the device fabrication). As the OEO material, we employed the side-chain EO polymer[71] with a glass transition temperature of 132.5 °C. Ti/Ni/Au electrode with ground-signal-ground (GSG) pads were integrated to enable high-speed characterization. The active metasurface region had a footprint of $40 \times 40$ µm$^2$. The electron mobility and doping concentration of the n-InP membrane layer were measured to be 1670 cm$^2$/Vs and $3.4 \times 10^{18}$ cm$^{-3}$, respectively, which are consistent with the values reported in the literature. The electrical capacitance of the device was 227 fF, which also agrees well with the FEM analysis (see Supplementary Note 3).

Figure 4a shows the measured reflectance and transmittance spectra of the IM-HCG with $\Lambda = 750$ nm and $w = 405$ nm. The experimental setup is depicted in Fig. 4b (see **Methods** for details of characterization). We can see a clear optical resonance at $\lambda = 1510$ nm with a $Q$ factor of 102. We can confirm that the sum of the reflectance $R$ and transmittance $T$ is more than 0.88 ($-0.56$ dB) over the entire wavelength range, indicating that our device is almost lossless. By applying a direct current (DC) voltage to the device, a clear shift of resonance wavelength at an efficiency of 18 pm/V is observed through the Pockels effect. Figure 4c shows the reflected and transmitted optical power observed when we apply 20-$V_{pp}$ sinusoidal voltage signal to the device. We can see that the transmittance and reflectance are complementarily modulated at the opposite phases. The modulation depths of the reflectance and transmittance are $\Delta R/R = 3.1\%$ and $\Delta T/T = 3.35\%$, respectively.

Finally, Fig. 5a shows the frequency response of the device, measured using the setup shown in Fig. 5b. The device exhibits a 3-dB bandwidth of 17.5 GHz, which agrees excellently with the FEM analysis (see Fig. 2g). To the best of our knowledge, this represents the largest 3-dB bandwidth among all experimentally demonstrated active metasurfaces reported to date (see Supplementary Table 1).

## Discussion

We have demonstrated an ultra-high-speed IM-HCG active metasurface device. Thanks to the large electron mobility of n-InP, our device exhibits a record-high 3-dB EO bandwidth of 17.5 GHz. This is nearly 6-fold improvement from the previously demonstrated active metasurfaces of all kinds, including plasmonic metasurfaces with metallic electrodes (see Supplementary Table 1 for the comparison). At the same time, owing to the small free-carrier absorption of n-InP, our device efficiently traps the incident light within a thin OEO layer without incurring significant optical absorption, so that $R + T > 0.88$, which corresponds to an ultra-low optical loss of 0.56 dB. This is also the record-low value among the active metasurfaces reported in the literature, while it exhibits a decently large $Q$ factor, exceeding 100.

The demonstrated device has a plenty of room for improvement in terms of modulation efficiency. First, the measured $Q$ factor of the resonator is ∼ 102, which is substantially smaller than that ($Q \sim 900$) expected from numerical simulation. This inconsistency is attributed to the imperfections in forming perfect IM-HCG structure, such as the sidewall roughness (Figs. 3c,d), which can be improved by optimizing the fabrication conditions. In addition, the EO coefficient $r_{33}$ of our fabricated device is estimated to be ∼ 20 pm/V, assuming $r_{33} \sim 3r_{13}$[72]. We expect that $r_{33}$ can be increased by employing an OEO material with a higher $r_{33}$ coefficient (as large as 500 pm/V[70]) and by improving the poling conditions. Assuming that the $Q$ factor and the $r_{33}$ coefficient are enhanced to 930 and 200 pm/V, respectively, we numerically estimate that a 5-dB modulation can be achieved with a driving voltage of approximately 5.6 $V_{pp}$.

Furthermore, the EO modulation bandwidth can be extended by increasing the doping concentration of n-InP. While the actual doping concentration of n-InP in our fabricated device was approximately $3.4 \times 10^{18}$ cm$^{-3}$, we have numerically demonstrated that it could be increased to $1 \times 10^{19}$ cm$^{-3}$ without inducing a substantial optical loss (Fig. 2c). Additionally, reducing the device size should be effective in extending the modulation bandwidth further due to smaller overall capacitance. We numerically estimate that an ultra-broad bandwidth exceeding 100 GHz should be attained at a device size of $24 \times 24$ µm$^2$ (see Supplementary Fig. S7). Our work, therefore, provides the pathway towards the realization of ultra-high-speed spatiotemporal control of lightwaves using active metasurfaces.

## Methods

### Numerical analysis of the optical characteristics

The optical characteristics of the IM-HCG and SM-HCG shown in Figs. 1d and 2a-f were obtained by two-dimensional finite-difference time-domain (FDTD) simulation (Ansys Lumerical FDTD) in the $zx$ plane. For simplicity, the HCGs were assumed to be infinitely long in the $y$ direction and the periodic boundary condition was employed in the $x$ direction. A normally incident $y$-polarized plane wave was input to the HCG. The refractive indices at $\lambda = 1550$ nm were used for all materials and the chromatic dispersion of the materials was neglected for simplicity. The duty ratio of the grating was set to $w/\Lambda = 0.5$ for all cases.

The absorption coefficients of n-InP and n-Si for various levels of doping concentration were obtained by using the values from the literature[73–76] to fit an empirical expression of $\alpha(N_D) = A \cdot (N_D)^B$, where $A$ and $B$ are the fitting parameters. The effect of absorption was then taken into the FDTD simulation by modifying the imaginary part of the complex refractive index of either n-InP or n-Si. The doping-induced changes in the real part of the refractive index was neglected to avoid the shift in the resonant wavelength, allowing direct comparison among different doping conditions at a same wavelength. For the optical loss plotted in Fig. 2c, we selected the largest value across the entire wavelength range shown in Figs. 2a and 2b.

### Numerical analysis of the electrical characteristics

The electrical characteristics of the device was simulated using the three-dimensional finite element method (FEM) (Ansys Lumerical CHARGE). The doping-concentration-dependent mobility was obtained from the Caugh-Thomas-like model[77] for n-InP and from the Masetti model[78] for n-Si (see Supplementary Note 5 for details).

### Device fabrication

The fabrication flow is shown in Supplementary Fig. S2. First, an InP epitaxial substrate [n-InP (590 nm), InGaAs (50 nm), InP (50 nm), InGaAs (100 nm), InP (substrate)] was prepared by the metal organic vapor phase epitaxy (MOVPE) growth. Thin Al$_2$O$_3$ adhesive layers were formed on the InP epitaxial wafer and the quartz wafer by the atomic layer deposition (ALD) before they were bonded[79]. The bulk InP substrate and the sacrificial epitaxial layers were removed through the wet-etching processes using hydrochloric acid (for InP) and phosphoric acid (for InGaAs), leaving only the 590-nm-thick n-InP membrane on the quartz substrate. After sputtering SiO$_2$ as a hard mask layer, electron beam (EB) lithography was performed to define the IM-HCG pattern. The SiO$_2$ hard mask and InP were then etched by the reactive-ion etching (RIE) processes (CHF$_3$ and C$_4$F$_8$/SF$_6$ for SiO$_2$, cyclic process of CH$_4$/H$_2$ and O$_2$ for InP). After removal of the SiO$_2$ hard mask, Al$_2$O$_3$ passivation layer is formed by ALD. Subsequently, Ni/Ti/Au electrodes were formed through EB lithography, EB evaporation, and lift-off process. Then, the EO polymer is deposited by spin coating process, followed by the sputtering of SiO$_2$ layer to protect the EO polymer. The IM-HCG region is defined by photolithography, and RIE is performed to etch the SiO$_2$ (CHF$_3$) and EO polymer (CHF$_3$/O$_2$). Finally, the EO polymer was poled by applying an electric field of 73 V/µm to the EO polymer at an elevated temperature of 131 °C, followed by rapid cooling while maintaining the poling field.

### Electrical characterization

The sheet resistance and contact resistance of the device were characterized by the transfer length method (TLM). The resistivity of the 590-nm-thick n-InP membrane was measured to be $1.1 \times 10^{-3}$ Ω·cm. The contact resistance of the n-InP was as low as $9.5 \times 10^{-6}$ Ω·cm$^2$. The electron mobility and the carrier concentration of n-InP were measured using the Hall measurement method. The capacitance of the device was measured using a semiconductor parameter analyzer. See Supplementary Note 4 for further details of the test patterns and the characterization result.

**Electro-optic characterization**

The DC and low-frequency measurement of the EO characteristics of the device was performed using the setup depicted in Fig. 4b. Continuous-wave (CW) light was launched from the tunable laser diode (TLD), controlled to *y* polarization, and emitted to free space via a fiber collimator. The light was then delivered to the IM-HCG device via a 3-dB beam splitter (BS) and an objective lens. The reflected light was collected by the same objective lens and coupled to a fiber via the beam splitter and a fiber collimator. The reflected optical power was measured by a power meter (PM) for the reflectance spectrum measurement and by an optical receiver consisting of an avalanche photodiode (APD) and trans-impedance amplifier (TIA) for the low-frequency modulation measurement. The transmitted light was measured by a bucket power meter. The reflectance and transmittance were calibrated by normalizing them to the measured values of reflection from a reference gold mirror and transmittance without any device, respectively. The voltage signal was applied to the device via DC probes. We used a picoammeter (PA) for applying a DC voltage and a function generator for applying a sinusoidal voltage. The Q factor of the device was extracted by fitting the reflection spectra to the Fano resonance model (see Supplementary Note 6).

The high-speed characterization of the device was carried out using the setup shown in Fig. 5b. The fiber-coupled reflected light was amplified by an erbium-doped fiber amplifier (EDFA), filtered by an optical band pass filter (OBPF), power-controlled by a variable optical attenuator (VOA), and detected by a high-speed photodiode. The frequency response of the device was measured using a vector network analyzer (VNA). The signal launched from the VNA was applied to the device via an RF probe and the electrical signal from the photodiode was delivered to the VNA.

## Data availability

The data that support the findings in this study are included in the article and its supplementary information. Other data are available from the corresponding authors upon reasonable request.

## Acknowledgements

This work was supported in part by the commissioned research (JPJ012368C03601, JPJ012368C08801) by National Institute of Information and Communications Technology (NICT) and Japan Society for the Promotion of Science (JSPS KAKENHI JP23H05444, JP23H00272, JP21J11982, JP21J10272). T.F. and K.S. acknowledge supports from Materials Education program for the future leaders in Research, Industry, and Technology (MERIT) of the University of Tokyo. Portion of the device fabrication was conducted at the cleanroom facilities of d.lab in the University of Tokyo, supported by MEXT Nanotechnology Platform, Japan. The authors acknowledge Shingo Kaneta-Takada and Makoto Fujiwara for their technical supports.

## Author contributions

T.F. and T.T. conceived the concept. T.F. performed the numerical simulation with some assistance from H.M. and G.S.. K.S. and T.F. conceived the device fabrication flow. W.Y. and T.F. performed the epitaxial wafer growth with assistance from E.K.. T.F. and K.S. fabricated the device with assistance from H.M., E.K., G.S., R.T, and J.L.. K.S. and T.F. performed the electrical characterization. T.F. conceived and performed the optical and electro-optic characterization with some assistance from H.M. and G.S.. T.Y. and A.O. developed the EO polymer. K.T., M.T., S.T., and Y.N contributed to overall discussion and provided experiment facilities. T.T. supervised the project. T.F. and T.T wrote the manuscript with the inputs from all authors.

## Competing interests

The authors declare no competing interests.

## Additional information

**Supplementary information** The online version of this paper contains supplementary material.
**Correspondance** and requests for materials should be addressed to Taichiro Fukui and Takuo Tanemura.

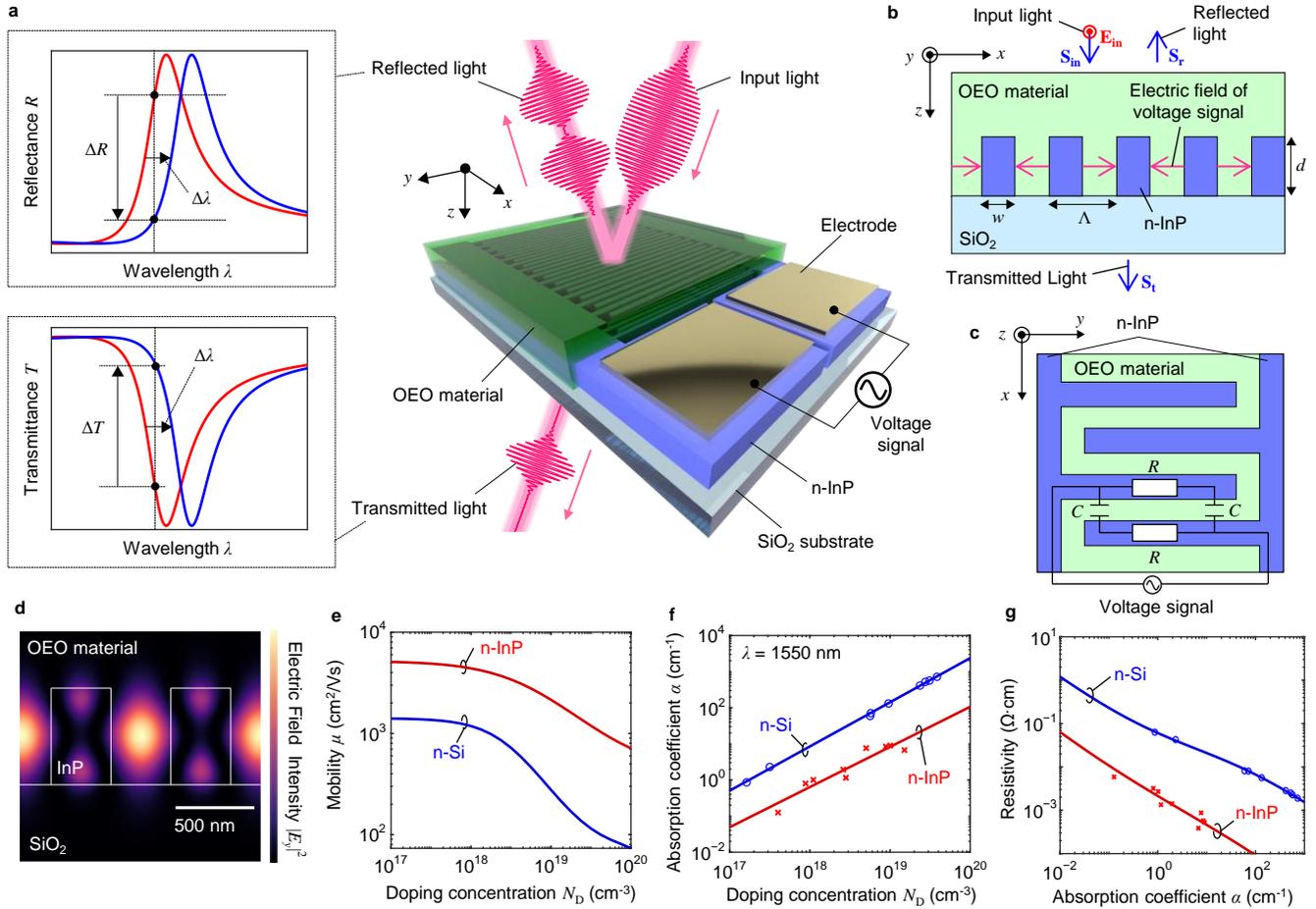

**Figure 1. Concept of the proposed InP-membrane active metasurface. a** Schematic of the proposed device. IM-HCG is realized by forming periodic InP-membrane bars embedded with OEO material on quartz substrate. InP bars simultaneously function as a resonator to trap the normally incident input light, and as an interdigitated electrode to apply voltage to the OEO material. By shifting the resonance wavelength, both the reflectance and transmittance of the IM-HCG can be modulated. **b** Cross-sectional schematic of IM-HCG. A normally incident $y$-polarized light is input to the device and couples to a resonating mode. By applying electric field to the EO polymer via the InP bars, the resonance wavelength is shifted. **c** Top-view schematic of IM-HCG. The InP bars form interdigitated electrodes to apply voltage signal to the OEO material. The modulation bandwidth of the device is limited by the RC time constant of the circuit, and thus the reduction of the resistivity of the InP bars is essential for high speed operation. **d** The electric field intensity ($|E_y|^2$) distribution of the resonating mode of the IM-HCG ($\lambda$ = 1550 nm, $\Lambda$ = 760 nm, $w$ = 380 nm, $d$ = 630 nm). The light is strongly confined in the OEO material, enabling efficient modulation. **e,f** Doping concentration dependence of the electron mobility (**e**) and absorption coefficient (**f**) for n-InP[73–75,77] and n-Si[76,78]. n-InP is superior to n-Si in both mobility and absorption across the entire range of doping concentration. **g** The tradeoff between the electrical resistivity and optical absorption for n-InP and n-Si.

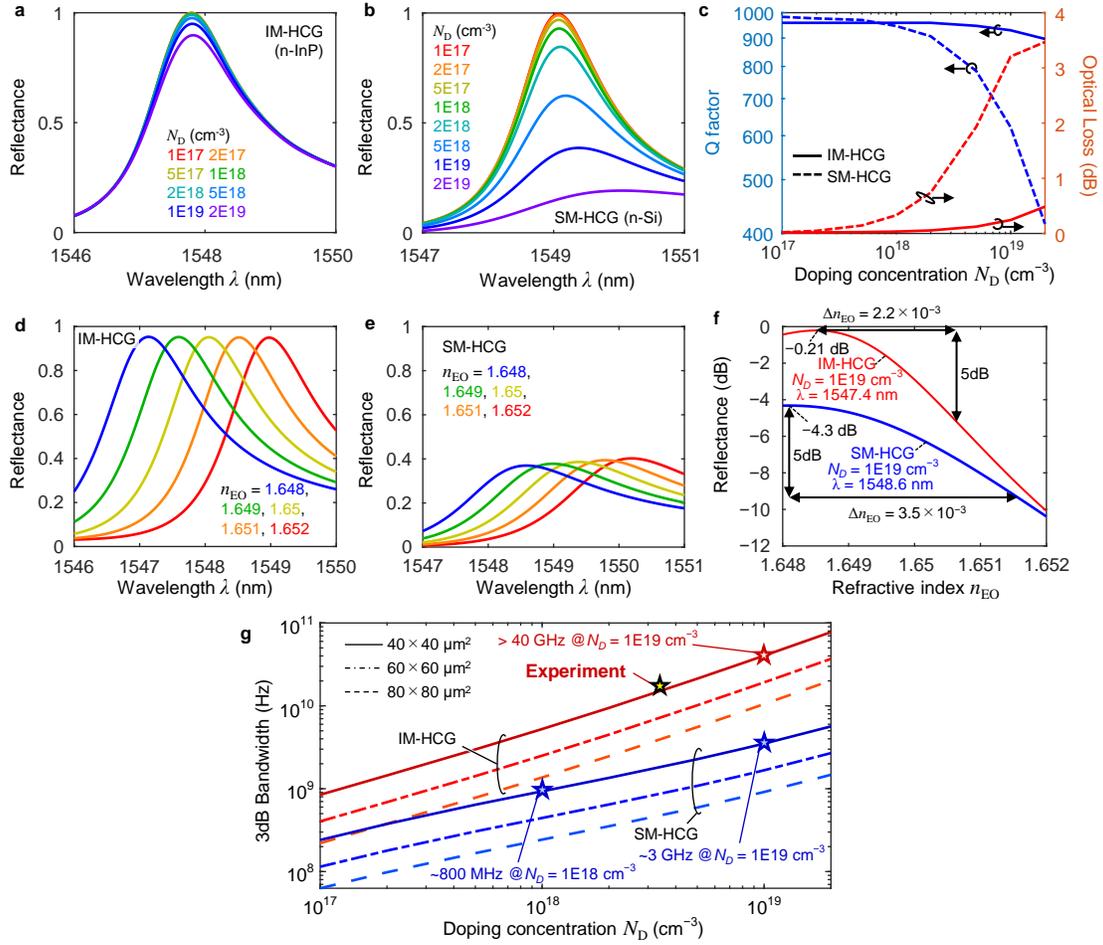

**Figure 2. Numerical analysis. a,b** The reflectance spectra of IM-HCG (**a**) and SM-HCG (**b**). **c** The $Q$ factor and optical loss at the resonance as a function of doping concentration for both IM-HCG and SM-HCG. **d,e** The reflectance spectra of IM-HCG (**d**) and SM-HCG (**e**) at $N_D = 1 \times 10^{19}$ cm$^{-3}$ for various $n_{EO}$. **f** Characteristics of IM-HCG and SM-HCG ($N_D = 1 \times 10^{19}$ cm$^{-3}$) when used as reflective modulators. **g** The 3-dB EO bandwidth of IM-HCG and SM-HCG active metasurfaces as a function of doping concentration.

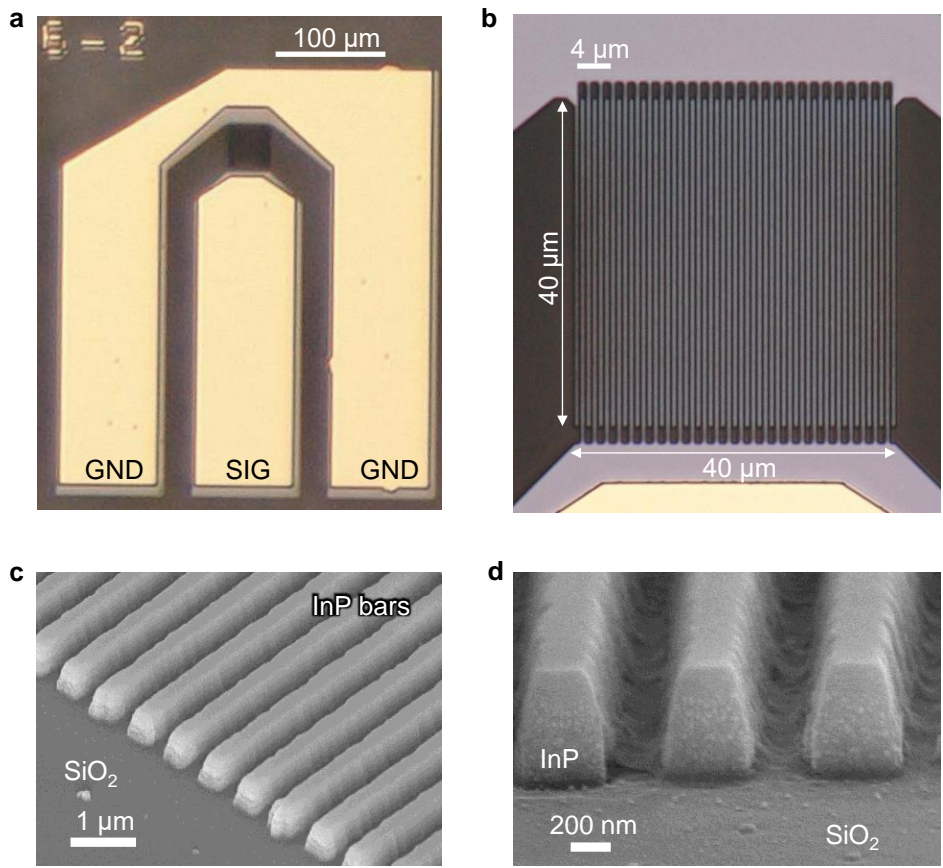

**Figure 3. Fabricated device. a** Microscopic image of the fabricated device. **b** Magnified view of the IM-HCG before coating the EO polymer. **c,d** Scanning electron microscope (SEM) images of the InP bars before coating the EO polymer.

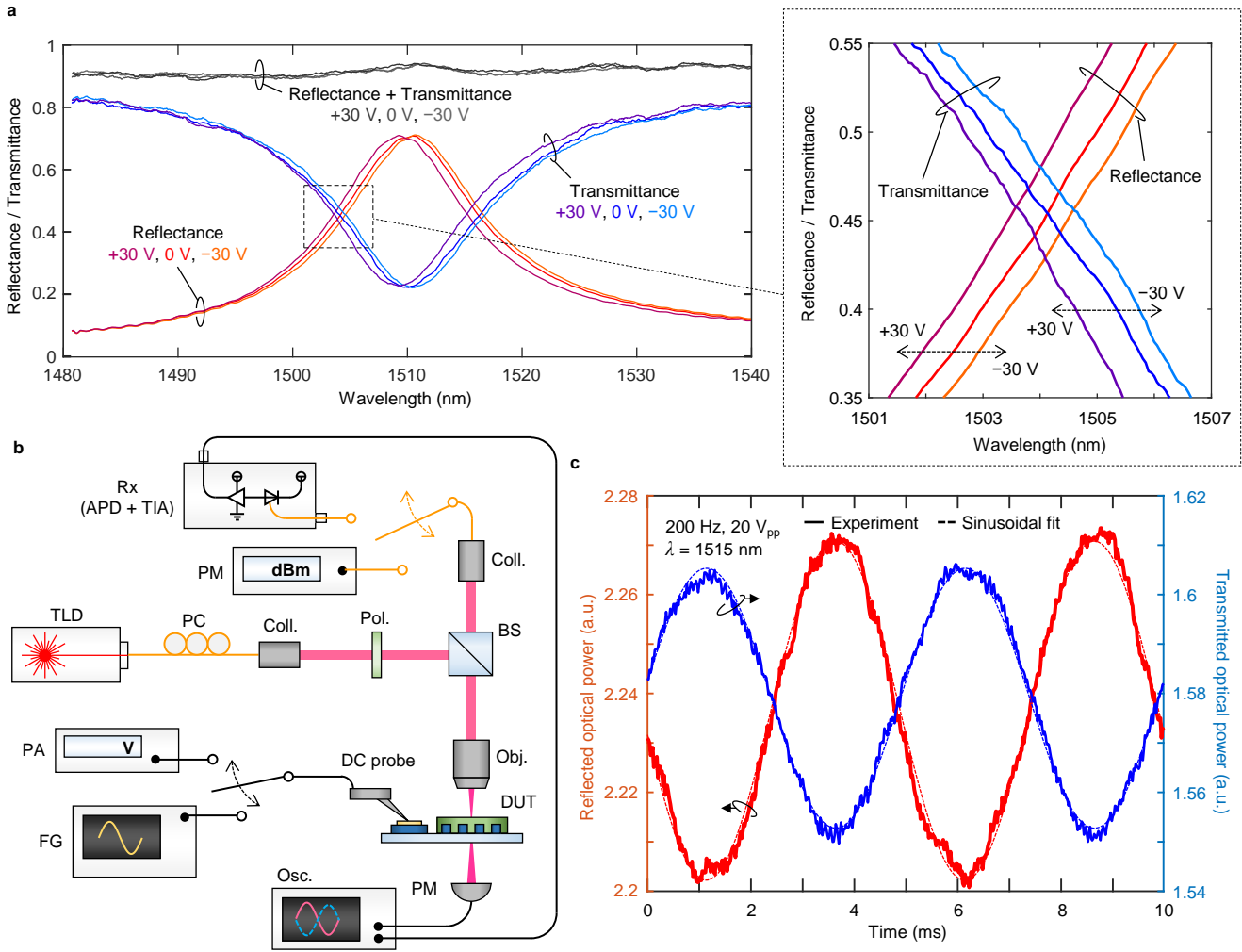

**Figure 4. DC and low-frequency characterization of the fabricated device. a** The reflectance ($R$) and transmittance ($T$) spectra of the device under different voltages (+30 V, 0 V, −30 V). The sum of $R$ and $T$ is also plotted, which exceeds 0.88 across the entire measured wavelength range. The magnified view is depicted in the right inset, which shows clear spectral shift under modulation. **b** Experimental setup. TLD: tunable laser diode. PC: polarization controller. Coll.: collimator. Pol.: polarizer. BS: beam splitter. Obj.: objective lens. DUT: device under test. PA: picoammeter. FG: function generator. PM: power meter. Rx: receiver. APD: avalanche photodiode. TIA: trans-impedance amplifier. Osc.: oscilloscope. **c** The reflected (red) and transmitted (blue) optical power measured when a 200-Hz 20-$V_{pp}$ sinusoidal voltage is applied to the device.

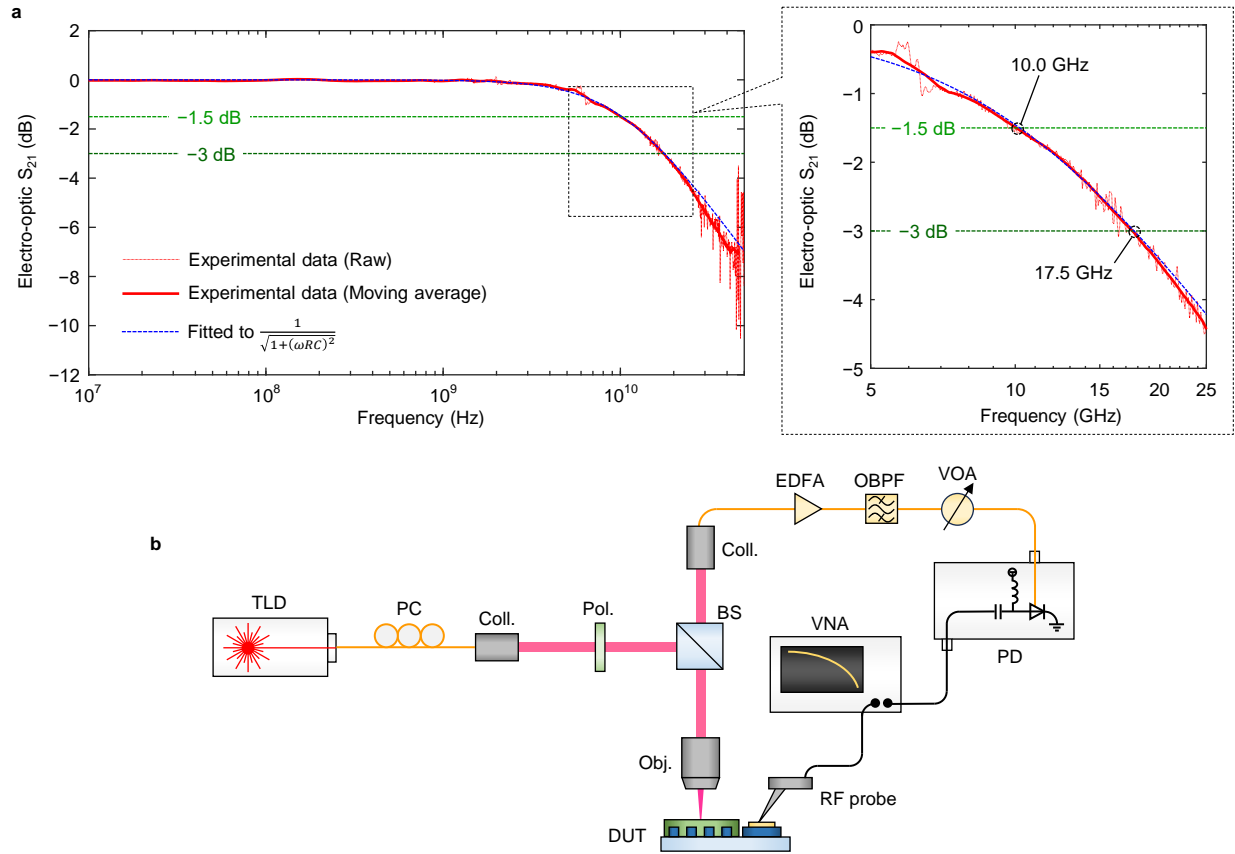

**Figure 5. High-speed characterization of the fabricated device. a** Measured EO frequency response of the device. A magnified view is depicted in the right inset. The 3-dB bandwidth is 17.5 GHz. **b** Experimental setup. The reflected light from the device is coupled to a fiber, amplified, and converted to an electrical signal by a biased photodiode (PD). VNA: vector network analyzer. EDFA: erbium-doped fiber amplifier. OBPF: optical bandpass filter.

# Supplementary Material for:
# 17 GHz Lossless InP-Membrane Active Metasurface

Taichiro Fukui[1,*], Kei Sumita[1], Hiroki Miyano[1], Go Soma[1], Warakorn Yanwachirakul[1,†], Eisaku Kato[1], Ryota Tanomura[1], Jiahao Liu[1], Toshiki Yamada[2], Akira Otomo[2], Kasidit Toprasertpong[1], Mitsuru Takenaka[1], Shinichi Takagi[1], Yoshiaki Nakano[1], and Takuo Tanemura[1,*]

[1]Department of Electrical Engineering and Information Systems, The University of Tokyo, 7-3-1 Hongo, Bunkyo-ku, Tokyo 113-8656, Japan
[2]National Institute of Information and Communications Technology, 588-2 Iwaoka, Nishi-ku, Kobe, Hyogo 651-2492, Japan
[†]Present address: Department of Physics, Faculty of Science, Chulalongkorn University, 254 Phayathai Road, Patumwan, Bangkok, Thailand, 10330.
[*]fukui@hotaka.t.u-tokyo.ac.jp; tanemura@ee.t.u-tokyo.ac.jp

## Supplementary Note 1: Bimodal resonance in HCG

Figure S1 shows the reflectance contour map of the IM-HCG as a function of the wavelength $\lambda$ and the thickness of the grating $d$ for a normal-incident $y$-polarized light, having the electric field oriented parallel to the grating. The inset shows the magnified view near the high-$Q$ resonant mode employed in our device. In these plots, we set $\Lambda = 760$ nm and the width of the InP bars to $w = 380$ nm, so that the duty ratio is $w/\Lambda = 0.5$. We assume that the refractive indices of all materials are constant: 3.17, 1.65, and 1.44 for InP, OEO material, and $SiO_2$, respectively. In the dual transverse mode region, high-$Q$ resonance can be achieved by judiciously tuning the thickness of the grating $d$[1–3]. Here, the transverse mode refers to the optical mode that propagates in the $z$-direction through an infinitely thick grating, subject to a periodic boundary condition in the $x$-direction.

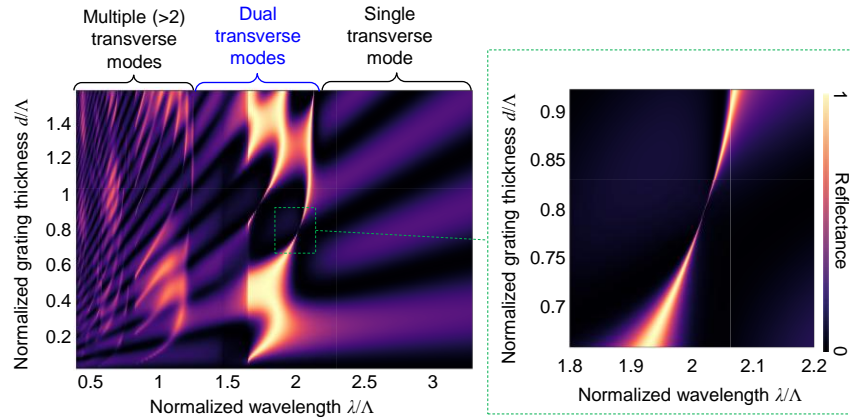

**Figure S1. Resonance spectrum of IM-HCG.** The reflectance contour map of the IM-HCG as a function of wavelength $\lambda$ and grating thickness $d$ normalized by the grating period $\Lambda$ ($\Lambda = 760$ nm, $w = 380$ nm). By judiciously designing $\Lambda$ and $d$, a sharp resonance can be obtained through the bimodal resonance effect. The inset shows the magnified view of the contour map near the resonant mode employed in our device.

## Supplementary Note 2: Fabrication process

Figure S2 shows the fabrication flow of the device. Details of the flow are described in the **Methods** section.

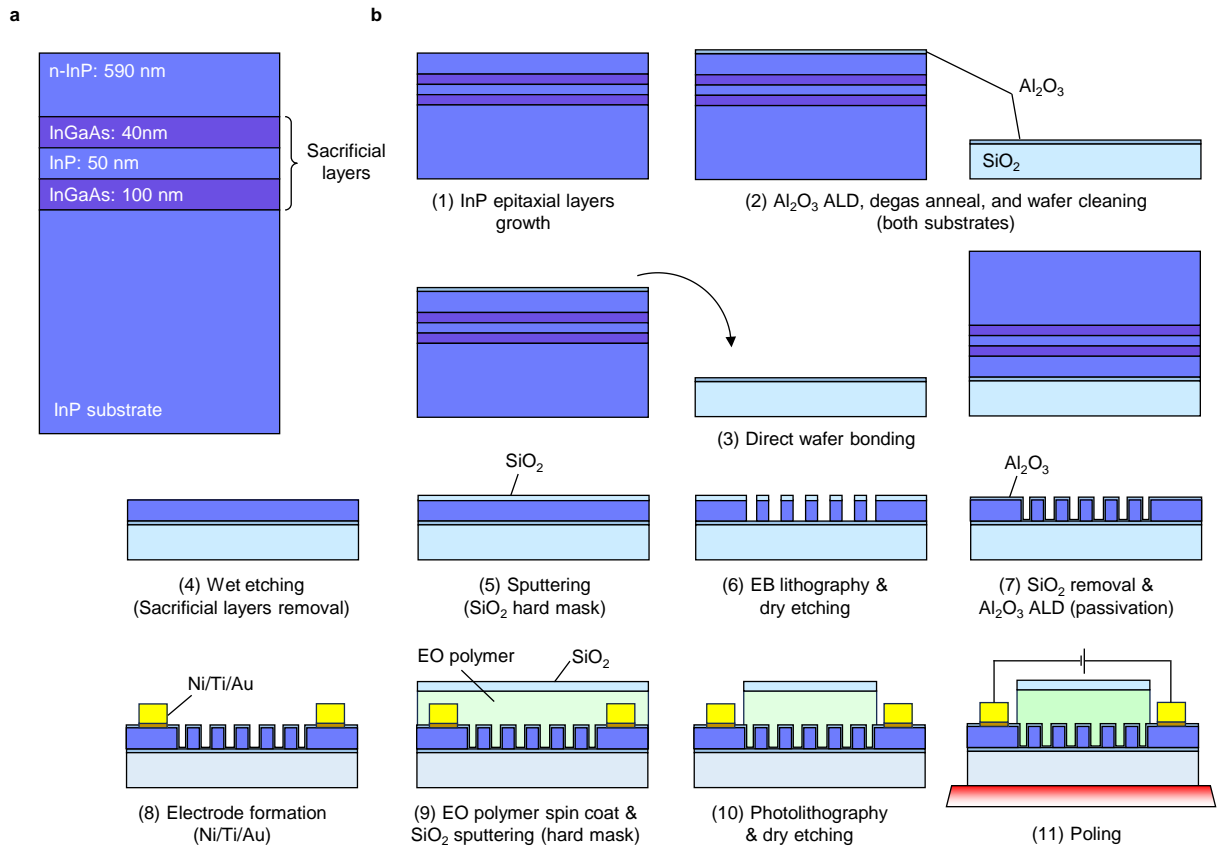

**Figure S2. Fabrication flow of the device. a** Epitaxial layers grown on InP substrate. **b** The fabrication flow of the device.

## Supplementary Note 3: Electrical capacitance of the device

Figure S3a shows the CV measurement result (measurement frequency: 1 MHz) of the devices with two different sizes, $40\times40$ µm$^2$ and $80\times80$ µm$^2$. The capacitance of the device is almost independent on the bias voltage. Figure S3b shows the simulated capacitance of the device for various size using finite element methods (FEM). Here, the doping concentration is set to $N_\text{D} = 3.4\times10^{18}$ cm$^{-3}$. The measured capacitance of the device shows good agreement with the numerical analysis, where the capacitance scales quadratically with the device size $L$. Figure S3c shows the simulated frequency-dependent impedance of the device. At low frequency, the lumped capacitance model ($|Z| = 1/\omega C$) describes the impedance characteristics well. At higher frequency, the impedance saturates, limited by the resistivity.

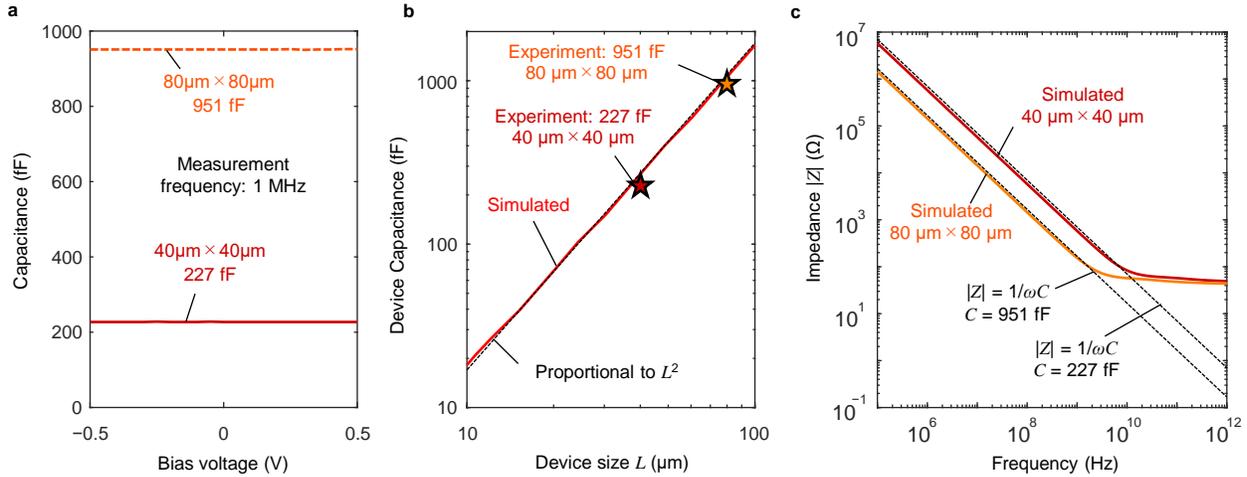

**Figure S3. Capacitance of the device. a** Experimentally measured capacitance-voltage characteristics of the fabricated devices with two different sizes, $40\times40$ µm$^2$ and $80\times80$ µm$^2$. **b** Simulated capacitance of the device as a function of size, $L$. Experimental results are also plotted, showing excellent agreements. **c** Simulated frequency-dependent impedance of the device.

## Supplementary Note 4: Electrical characterization of n-InP membrane

Figure S4a shows the schematic of the test pattern for the transfer length method (TLM). By measuring the resistivity with different gap length $L_g$ while keeping the electrode pattern same (width $W$), the sheet resistance $R_s$ of the InP membrane and the contact resistance $R_c$ of the Ni/Ti/Au contact to InP can be extracted. Figure S4b shows the TLM result. By fitting the data to the equation

$$R = 2R_c + R_s L_g / W, \tag{S1}$$

$R_s$ and $R_c$ are extracted to be 18.9 $\Omega\square$ and 1.68 $\Omega$, respectively. Given the InP membrane thickness $d = 590$ nm, the resistivity $\rho_s$ of InP is given as $\rho_s = R_s \times d = 1.1 \times 10^{-3}$ $\Omega\cdot$cm. On the other hand, according to the transmission line model, the contact resistivity $\rho_c$ of n-InP/Ni/Ti/Au contact can be derived as[4]

$$\rho_c = \frac{R_c^2 W^2}{R_s} = 9.5 \times 10^{-6} \ \Omega\cdot\text{cm}^2 \ . \tag{S2}$$

This small contact resistivity is attributed to the formation of InP/Ni alloy.[5]

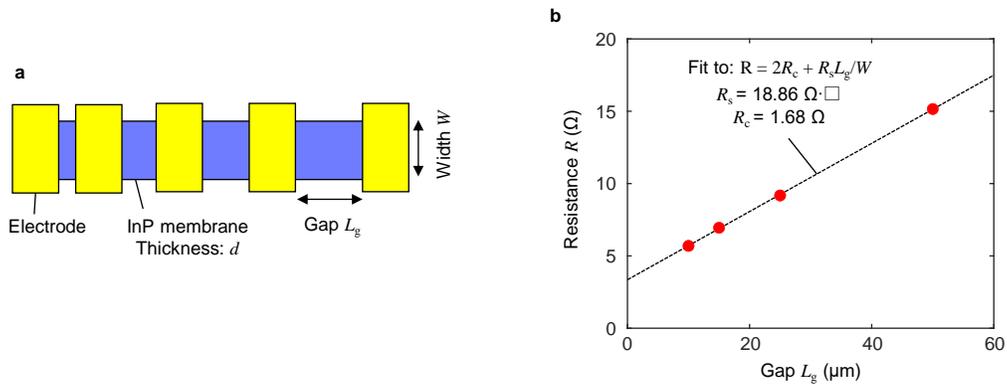

**Figure S4. Transmission line measurement (TLM). a** Schematic of the test pattern. **b** Measured results.

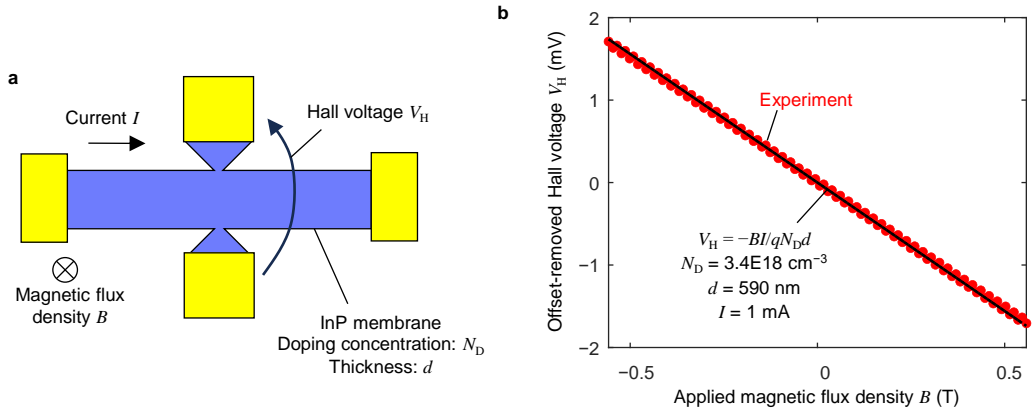

**Figure S5. Hall measurement. a** Schematic of the test pattern. **b** Measured results at $I = 1\,\text{mA}$.

Figure S5**a** shows the schematic of the test pattern for the Hall measurement. By measuring the Hall voltage $V_\text{H}$ under a constant injection current $I$ while varying the applied magnetic flux $B$, the carrier concentration can be extracted. Figure S5**b** shows the Hall measurement result (offset voltage is removed for clarity). $V_\text{H}$ varies linearly with $B$. By fitting the data to the equation

$$V_\text{H} = -\frac{BI}{qN_\text{D}d}, \tag{S3}$$

$N_\text{D}$ is derived as $3.4 \times 10^{18}$ cm$^{-3}$. The electron mobility $\mu$ is then derived as $\mu = 1/(q\rho_\text{s}N_\text{D}) = 1670$ cm$^2/\text{V}\cdot\text{s}$.

## Supplementary Note 5: Electron mobility model

In order to calculate the doping-concentration-dependent electron mobility of n-InP and n-Si, we employed empirical models. For n-InP, we used the Caughey-Thomas-like model, where the electron mobility is expressed as[6]

$$\mu(N_D) = \mu_{\min} + \frac{\mu_{\max} - \mu_{\min}}{1 + \left(\dfrac{N_D}{N_0}\right)^{0.47}}. \tag{S4}$$

Here, $N_0 = 3 \times 10^{17}$ cm$^{-3}$, $\mu_{\max} = 5200$ cm$^2$/V·s, and $\mu_{\min} = 420$ cm$^2$/V·s. Figure S6 shows the doping concentration-dependent electron mobility of n-InP obtained from the model and compares it with the experimentally obtained value. Good agreement between the model and the experiment is obtained.

For n-Si, we used the Masetti model[7], where the electron mobility is expressed as

$$\mu(N_D) = \mu_{\min} + \frac{\mu_{\max} - \mu_{\min}}{1 + \left(\dfrac{N_D}{C_r}\right)^{\alpha}} - \frac{\mu_1}{1 + \left(\dfrac{N_D}{C_s}\right)^{\beta}}. \tag{S5}$$

Here, $C_r = 9.68 \times 10^{16}$ cm$^{-3}$, $C_s = 3.43 \times 10^{20}$ cm$^{-3}$, $\alpha = 0.68$, $\beta = 2$, $\mu_{\max} = 1471$ cm$^2$/V·s, $\mu_{\min} = 52.2$ cm$^2$/V·s, and $\mu_1 = 43.4$ cm$^2$/V·s.

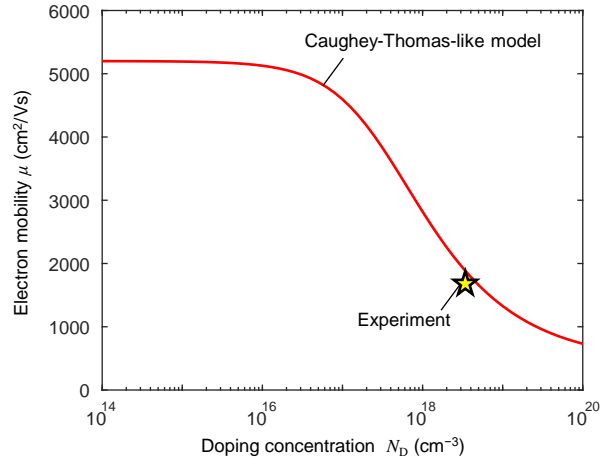

**Figure S6. Electron mobility of n-InP.** Doping concentration dependence of the electron mobility of n-InP based on the Caughey-Thomas model. The experimentally obtained value agrees well with the model.

## Supplementary Note 6: Bandwidth simulation

Figure S7 shows the simulated 3-dB modulation bandwidth as a function of device size $L$. At a device size of $24\,\mu\text{m} \times 24\,\mu\text{m}$, the 3-dB modulation bandwidth exceeds 100 GHz. Note that the 3 dB bandwidth roughly scales with $1/L^2$, where $L$ denotes the device size (or the bar length). This is because both the resistivity and capacitance of each InP bar scale linearly with $L$. The slight deviation from the $1/L^2$ trend is attributed to the effect at the edges of the device.

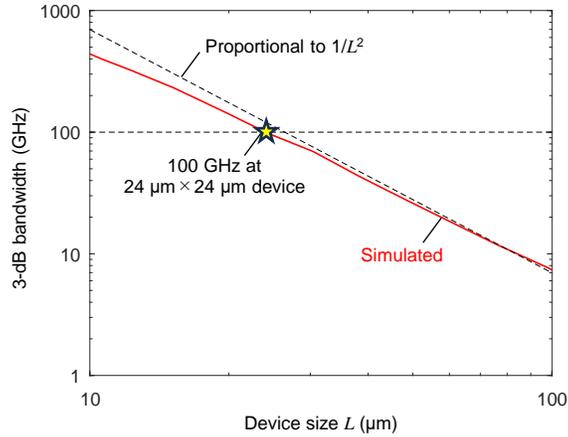

**Figure S7. Size dependence of the modulation bandwidth.** Simulated 3-dB modulation bandwidth of the device (with $N_\text{D} = 1 \times 10^{19}\,\text{cm}^{-3}$) as a function of the device size $L$.

## Supplementary Note 7: Fano resonance fitting

To determine the $Q$ factor and resonance wavelength of our device, we fitted the measured reflectance spectrum to the Fano resonance model, described as[8]

$$R(\omega) = \left| r_d + \frac{a_1 \gamma_1}{\gamma_1 + i(\omega - \omega_1)} + \frac{a_2 \gamma_2}{\gamma_2 + i(\omega - \omega_2)} \right|^2, \tag{S6}$$

where the first term $r_d$ denotes the direct reflection, the second term corresponds to the high-$Q$ resonance, the third term corresponds to the low-$Q$ background resonance, $a_i$ denotes the amplitude of the i-th resonance, $\gamma_i$ denotes the decay rate of the i-th resonance, and $\omega_i$ denotes the resonance frequency of the i-th resonance, respectively. Note that $r_d$ is generally complex due to the phase change induced by direct reflection. Figure S8a shows the resonance spectrum of the device at 0 V with the fitting result. The experimental result shows good agreement with the Fano resonance model. Here, $|r_d| = 0.0456$, $\arg(r_d) = 0.7769$ rad, $a_1 = 0.6097$, $\gamma_1 = 6.0936 \times 10^{12}$ s$^{-1}$, $\omega_1 = 1.2474 \times 10^{15}$ rad/s, $a_2 = 0.2063$, $\gamma_2 = 3.5938 \times 10^{13}$ s$^{-1}$, and $\omega_2 = 1.240 \times 10^{15}$ rad/s.

From the fitting result, the $Q$ factor can be derived as[9]

$$Q = \frac{\omega_1}{2\gamma_1}. \tag{S7}$$

The extracted $Q$ factor is depicted in Fig. S8b, which is almost independent on the applied voltage. Figure S8c shows the extracted resonance wavelength $\lambda_1 = 2\pi c / \omega_1$ ($c$ is the speed of light) as a function of the applied voltage. The resonance wavelength shift per voltage is derived to be 18 pm/V.

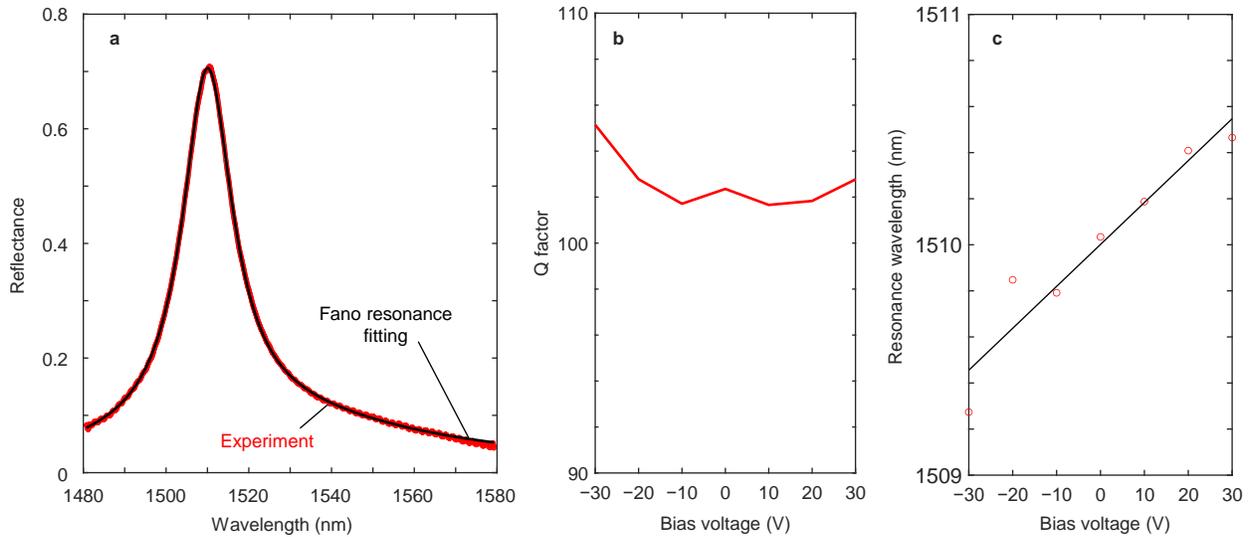

**Figure S8. Fitting the reflectance spectrum to Fano resonance model. a** Reflectance spetrum of the device fitted to the Fano resonance spectrum. **b** The $Q$ factor of the device as a function of applied DC voltage. **c** The resonance wavelength of the device as a function of applied DC voltage.

## Supplementary Note 8: Benchmark

Table S1 compares the performance metrics of our device with those of the active metasurfaces experimentally demonstrated in the literature.

Our device experimentally exhibits the record-high 3-dB bandwidth of 17.5 GHz, which is nearly a 6-fold improvement over previously demonstrated active metasurfaces of all kinds reported to date, including plasmonic metasurfaces with metallic electrodes. This is attributed to the low resistivity of n-InP and optimized device geometry to reduce the parasitic capacitance. The excess optical loss (0.56 dB) of our device is also a record low loss among various active metasurfaces reported to date, thanks to the small optical absorption of n-InP.

**Table S1. Comparison of performance metrics of active EO metasurfaces.** In the table, the active material, resonator material, and electrode material refer to the material whose refractive index is modulated, the nanostructured material that functions as the resonator, and the material used to apply signal to the active material, respectively. LN: $LiNbO_3$. BTO: $BaTiO_3$. ITO: indium tin oxide. PE: Pockels effect. FCE: free carrier effect. QCSE: quantum confined Stark effect.

| Active material | Modulation scheme | Resonator material | Electrode material | Wavelength (nm) | $Q$ factor | 3 dB bandwidth $f_{3dB}$ (GHz) | Optical loss[#] (dB) | Efficiency $\Delta\lambda/V$ (pm/V) | Reference |
|---|---|---|---|---|---|---|---|---|---|
| OEO | PE | InP | InP | 1510 | 102 | **17.5** | **0.56** | 18 | This work (Experiment) |
| OEO | PE | InP | InP | 1550 | 930 | **40** | **0.24** | 180 | This work (Numerical) |
| OEO | PE | Si | Si | 1510 | 75 | N/A[a] | N/A[†] | 16 | [2] |
| OEO | PE | Si | Si/Au | 1510 | 1310 | $3 \times 10^{-3}$ | N/A[†] | 105 | [10] |
| OEO | PE | Si | Au | 1540 | 550[b] | 3 | 4.1[c] | 55 | [11] |
| OEO | PE | Si | Au/ITO | 1310 | 153 | 0.4 | 11.4* | 22 | [12] |
| OEO | PE | Au | Au | 1400 | 70 | 0.05 | N/A[††] | 70 | [13] |
| OEO | PE | Au | Au/ITO | 1280 | 200 | 0.1 | 9.7* | 28 | [14] |
| OEO | PE | Au | Au | 1290 | 77 | 0.08 | 12.9* | 7.4 | [15] |
| OEO | PE | Au | Au | 1650 | 113 | 1.25 | 27* | 40 | [16] |
| LN | PE | LN | Au | 770 | 129 | $2.5 \times 10^{-3}$ | N/A[††] | N/A[‡] | [17] |
| LN | PE | Al | ITO/Al/Cr/Au | 900 | 30 | $7.78 \times 10^{-4}$ | 12.2* | 60 | [18] |
| LN | PE | Au | Au | 900 | 30 | $1.35 \times 10^{-2}$ | 17* | 60[‡‡] | [19] |
| LN | PE | Au | Au | 1550 | 180 | 0.125 | 6.2* | 40[‡‡] | [20] |
| BTO | PE | BTO | ITO | 630 | 200 | $2 \times 10^{-3}$ | N/A[††] | 50[‡‡] | [21] |
| Si | FCE | Si | Si | 1550 | 3500[d] | N/A[e] | N/A[f] | 750[g] | [22] |
| InGaAsP | QCSE | AlGaAs | Au | 920 | N/A | N/A[h] | 12.1 | 890 | [23] |

[#] Defined as $-10\log_{10}(R+T)$, where $R$ and $T$ denotes the reflectance and transmittance, respectively.
[†] Transmission data is not available.
[††] Reflection data is not available.
[‡] Not reported.
[‡‡] Simulated.
* Designed to perfectly absorb the incident light to function as an intensity modulator.
[a] The frequency response is not reported but modulation at 30 MHz is reported.
[b] The value is extracted for low-speed device. The Q factor reported for the high-speed device is 90.
[c] Numerical result, where the simulated Q factor of device is 610.
[d] Degraded value due to absorption by injected free carrier. Without injection, $Q$ = 4300.
[e] The frequency response is not reported but modulation at 150 MHz is demonstrated.
[f] Both reflectance and transmittance spectra are reported but the absolute value is not reported.
[g] It is not necessarily appropriate to compare by wavelength shift per voltage as the device operates at the carrier injection mode.
[h] Modulation at 1 MHz is reported.